\documentclass[prd,aps,twocolumn,nofootinbib]{revtex4}

\usepackage{graphicx,amsmath}

\begin{document}

\title{Throat effects on shadows of Kerr-like wormholes}

\author{Shinta Kasuya and Masataka Kobayashi}

\affiliation{
Department of Mathematics and Physics,
     Kanagawa University, Kanagawa 259-1293, Japan}
     
\date{March 29, 2021}

\begin{abstract}
We revisit to investigate shadows cast by Kerr-like wormholes. The boundary of the shadow is determined
by unstable circular photon orbits. We find that, in certain parameter regions, the orbit is located at 
the throat of the Kerr-like wormhole, which was not considered in the literature. In these cases, 
the existence of the throat alters the shape of the shadow significantly, and makes it possible for us
to differentiate it from that of a Kerr black hole.
\end{abstract}

\maketitle

\section{Introduction}
A wormhole is a spacetime bridge connecting two separate points in spacetime 
\cite{Flamm,Einstein:1935tc,Ellis:1973yv,Morris:1988cz}. 
(For early review, see \cite{Visser:1995cc}, for example.) Astrophysically, it is plausible to regard 
that wormholes are rotating. One of the metric for the rotating wormhole is the Kerr-like wormhole 
proposed in Ref.~\cite{Bueno:2017hyj}. It is important to distinguish this wormhole from 
the Kerr black hole, and one such way is to study shadows of the Kerr-like wormholes \cite{Amir:2018pcu}.
Remarkable attention to shadows cast by compact objects grows recently, since the Event Horizon Telescope 
observed the black hole shadow at the center of the galaxy M87 \cite{Akiyama:2019cqa}.

In this article, we revisit to investigate shadows of the Kerr-like wormhole, studied previously in
Ref.~\cite{Amir:2018pcu}. The shadow boundary can be estimated by considering unstable circular 
photon orbits. We point out that the effects of the wormhole throat become prominent 
when the unstable circular photon orbits are located at the throat in certain parameter space,
which was overlooked before.
In these cases, the shape of the shadow is altered considerably, and thus we can figure out the differences
between the shadow shapes of the Kerr-like wormhole and the Kerr black hole. 
In the study of the shadows of other types of the rotating wormholes \cite{Nedkova:2013msa}, 
similar results were obtained in Ref.~\cite{others}, where the throat 
effects on the shadows were taken into account.

The structure of the article is as follows. In the next section, we introduce the metric of the
Kerr-like wormhole, and show some of its features. In Sec.~\ref{sec3}, we derive unstable circular
photon orbits from the radial geodesic equation, where we recollect the results in Ref.~\cite{Amir:2018pcu} 
in Sec~\ref{sec_A}, and point out that the unstable circular photon orbit can lie at the throat in some
parameter regions in Sec.~\ref{sec_throat}. We show the shapes of the shadows on the celestial plane 
with and without the effects of the throat in Sec.~\ref{sec_shadow}. Section \ref{concl} is devoted
to our conclusions. We set $G=c=1$ throughout the article.

\section{Kerr-like wormhole}
The metric of the Kerr-like wormhole was constructed with slight modification from the metric of the 
Kerr black hole, which reads, in Boyer-Lindquist coordinates, as \cite{Bueno:2017hyj,Amir:2018pcu}
\begin{align}
ds^2 = & -\left(1-\frac{2Mr}{\Sigma}\right)dt^2-\frac{4Mar\sin^2\theta}{\Sigma}dtd\phi \notag \\
& +\frac{\Sigma}{\hat{\Delta}}dr^2+\Sigma d\theta^2 \notag \\
& +\left(r^2+a^2+\frac{2Ma^2 r\sin^2\theta}{\Sigma}\right)\sin^2\theta d\phi^2,
\label{metric}
\end{align}
where
\begin{align}
\Sigma & = r^2 + a^2 \cos^2\theta, \\
\hat{\Delta} & = r^2 +a^2 - 2M(1+\lambda^2) r.
\label{Delta-hat}
\end{align}
Here $\lambda$ represents the deviation parameter such that the metric reduces to that 
of the Kerr black hole for $\lambda=0$. $M$ and $aM$ respectively denote the mass and 
the angular momentum of the would-be Kerr black hole. This metric is a rotational version 
of the Schwarzschild-like wormhole proposed in Ref.~\cite{Damour:2007ap}, 
which is recovered by setting $a=0$. 

Instead of having an event horizon, the Kerr-like wormhole spacetime has a throat. It is evaluated by 
$\hat{\Delta}(r_{\rm throat})=0$, so that we obtain the throat radius as~\cite{Bueno:2017hyj}
\begin{equation}
r_{\rm throat} = M(1+\lambda^2)+\sqrt{M^2(1+\lambda^2)^2-a^2}.
\end{equation}
The ADM mass of the Kerr-like wormhole is estimated as \cite{Amir:2018pcu}
\begin{equation}
M_{\rm WH}^{\rm (ADM)} = M(1+\lambda^2),
\label{ADM}
\end{equation}
as seen by an observer at the asymptotic spatial infinity.

\section{Unstable circular photon orbits}
\label{sec3}
Null geodesic equations govern the trajectory of light. For stationary and axisymmetric spacetime, 
it is characterized by three constants: the photon energy $E$, angular momentum $L$, and the 
Carter constant ${\cal K}$ \cite{Carter:1968rr}. Four null geodesic equations are obtained 
as \cite{Bueno:2017hyj, Amir:2018pcu}
\begin{align}
\Sigma\frac{dt}{d\sigma} & = -a(aE\sin^2\theta-L)+\frac{(r^2+a^2){\cal P}}{\Delta}, \\
\label{phi_eq}
\Sigma\frac{d\phi}{d\sigma} & = -\left(aE-\frac{L}{\sin^2\theta}\right)+\frac{a{\cal P}}{\Delta},\\
\label{r_eqn}
\Sigma\frac{dr}{d\sigma} & = \pm \sqrt{\cal R}, \\
\label{theta_eq}
\Sigma\frac{d\theta}{d\sigma} & = \pm \sqrt{\Theta},
\end{align}
where $\sigma$ is the affine parameter, and
\begin{align}
\label{Delta}
\Delta & = r^2+a^2-2Mr, \\
{\cal P} & = (r^2+a^2)E-aL, \\
{\cal R} & = \frac{\hat\Delta}{\Delta}\left\{ {\cal P}^2-\Delta\left[{\cal K}+(L-aE)^2\right]\right\}, \\
\Theta & = {\cal K} + \cos^2\theta\left( a^2E^2-\frac{L^2}{\sin^2\theta}\right).
\end{align}

We consider a wormhole connecting two regions of the spacetime, where light sources illuminate around
the wormhole in one region, and there is none in the vicinity of the wormhole in the other region. 
In the former region, the light propagation can be divided into two categories. The first one is that 
the light plunges into the wormhole passing its throat away to the other region, and the second one is that 
the light is scattered away to the infinity in the same region. The distant observer in this region
sees the bright and dark zones whose boundary is determined by that of the two categories.
This dark zone is referred to as the shadow of the wormhole. 

In order to find the shadow, therefore, we need to specify that boundary, which is represented 
by the unstable circular photon orbits around the wormhole. To this end, we rewrite the radial geodesic
equation (\ref{r_eqn}) as
\begin{equation}
\frac{1}{2}\left(\frac{dr}{d\sigma}\right)^2 + V_{\rm eff} = 0.
\end{equation}
Here $V_{\rm eff}(r)$ can be regarded as the effective potential, given by
\begin{equation}
V_{\rm eff} = - \frac{1}{2\Sigma^2}{\cal R}= - \frac{E^2}{2}\frac{\hat{\Delta}(r)X(r)}{Y(r)}
\equiv - \frac{E^2}{2} \tilde{V}_{\rm eff},
\end{equation}
where we use the impact parameters $\xi = L/E$ and $\eta={\cal K}/E^2$, and 
\begin{align}
X(r) & = (r^2+a^2-a\xi)^2-\Delta(r)\left[ \eta + (\xi - a)^2 \right], \\
Y(r) & = \Sigma^2(r) \Delta(r).
\end{align}

The radius of the unstable circular photon orbit $r_{\rm ph}$ is derived by the conditions:
\begin{equation}
V_{\rm eff}(r_{\rm ph}) = 0, \quad V_{\rm eff}'(r_{\rm ph}) = 0, \quad V_{\rm eff}''(r_{\rm ph}) < 0,
\end{equation}
where the prime denotes the derivative with respect to $r$, which leads to the conditions in terms of 
$\tilde{V}_{\rm eff}$ as 
\begin{equation}
\tilde{V}_{\rm eff}(r_{\rm ph}) = 0, \quad \tilde{V}_{\rm eff}'(r_{\rm ph}) = 0, 
\quad \tilde{V}_{\rm eff}''(r_{\rm ph}) > 0.
\label{cond2}
\end{equation}
Since $Y(r)\ne 0$ for $r> r_{\rm throat}$, the first condition is satisfied for (i) $X(r_{\rm ph}) =0$, 
or (ii) $\hat{\Delta}(r_{\rm ph})=0$. The case (i) was investigated in Ref.~\cite{Amir:2018pcu},
but the case (ii) was overlooked. We will come back to the case (ii) in Sec.~\ref{sec_throat}, and
keep considering only the case (i) in the following subsection. 

\subsection{Case (i) $X(r_{\rm ph}) =0$}
\label{sec_A}
In this case, the second and third conditions read as $X'(r_{\rm ph})=0$ and $X''(r_{\rm ph})>0$,
respectively. Then, from the first and second conditions, the impact parameters can be given 
in terms of $r_{\rm ph}$ as \cite{Amir:2018pcu}
\begin{equation}
\begin{split}
\xi & = \frac{r_{\rm ph}^2(r_{\rm ph}-3M)+a^2(r_{\rm ph}+M)}{a(M-r_{\rm ph})}, \\
\eta & = \frac{4a^2Mr_{\rm ph}^3-r^4_{\rm ph}(r_{\rm ph}-3M)^2}{a^2(M-r_{\rm ph})^2}.
\end{split}
\label{impact1}
\end{equation}
The range of the radius of the unstable circular photon orbit $r_{\rm ph}$ is estimated as 
\begin{equation}
r_{\rm ph}^{\rm (min)}<r_{\rm ph}<r_{\rm ph}^{\rm (max)}, 
\label{rph}
\end{equation}
where
\begin{equation}
r_{\rm ph}^{\rm (min/max)} = 2M\left\{ 1 + \cos\left[\frac{2}{3}\cos^{-1}\left((-/+)\frac{a}{M}\right)\right]\right\},
\end{equation}
which correspond to the prograde/retrograde orbits on the equatorial plane, respectively \cite{Bardeen:1972fi}.
Of course, they are roots of the equation $\eta(r)=0$.

From the third condition, $X''(r_{\rm ph})>0$, $r_{\rm ph}$ must obey $r_{\rm ph} > r_{\rm 3rd}$, where
$r_{\rm 3rd}(>0)$ is a root of 
\begin{equation}
r^3-3M r^2+3M^2r-Ma^2=0.
\end{equation}
Since one can easily see that $r_{\rm 3rd}<M$, we always have $r_{\rm ph} > r_{\rm 3rd}$. 

The non-rotating case ($a=0$) should be considered separately. In this case, we obtain the fixed unstable 
circular orbit and the relation between two impact parameters respectively as
\begin{equation}
r_{\rm ph} = 3M, \qquad \eta = 27M^2 -\xi^2. 
\label{impact2}
\end{equation}
As for the third condition in Eq.(\ref{cond2}), $\eta$ and $\xi$ must satisfy the constraint
\begin{equation}
\eta < 54 M^2 -\xi^2,
\label{cond3-0}
\end{equation}
where we use $r_{\rm ph} = 3M$. Here this condition is always accomplished.

Notice that the expression of the unstable circular photon orbits, (\ref{impact1}) 
with (\ref{rph}), or (\ref{impact2}), are exactly the same as those for the Kerr black hole 
whose metric is given by Eq.(\ref{metric}) with $\lambda=0$. In Ref.~\cite{Amir:2018pcu}, 
the authors used the ADM mass as the mass of the Ker-like wormhole. 
Thus, when they compared the Kerr-like wormhole with the Kerr black hole with the same ADM mass,
they found that the radius of the unstable circular photon orbits is smaller for the Kerr-like 
wormhole than that of the Kerr black, since $M$ scales as $M_{\rm WH}^{\rm (ADM)}/(1+\lambda^2)$
for the Kerr-like wormhole, while the ADM mass of the Kerr black hole reads as $M_{\rm BH}^{\rm (ADM)}=M$.
\footnote{
We show the results using ADM masses in Appendix.}

However, it may be more appropriate to compare the shadow shapes of the two for the same mass determined by,
for example, the observations of dynamics of other surrounding objects. In this case, the mass of the
wormhole/black hole should be $m$, which causes almost Newtonian gravity, estimated from  
$-g_{tt} \approx 1-\frac{2m}{r}$.\footnote{
We thank the anonymous referee for pointing this out.}
This leads to the dynamically determined masses as $M_{\rm WH}^{\rm (dyn)}=M_{\rm BH}^{\rm (dyn)}=M$ in
our situations. Therefore, the shadow shapes of the Kerr-like wormhole and the Kerr black hole are
exactly the same if we do not take into account the throat effects.

\subsection{Case (ii) $\hat{\Delta}(r_{\rm ph})=0$}
\label{sec_throat}
\begin{figure}[ht!]
\includegraphics[width=90mm]{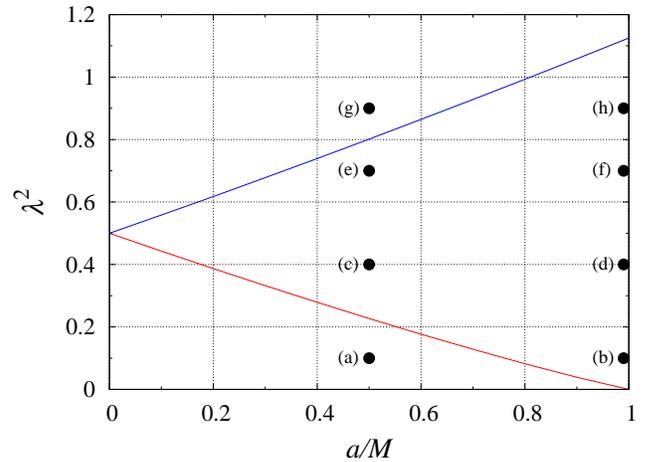} 
\caption{Regions where the throat affects the shape of shadow in parameter space $(a/M, \lambda^2)$.
Red and blue lines denote $r_{\rm throat}=r_{\rm ph}^{\rm (min)}$ and 
$r_{\rm throat} = r_{\rm ph}^{\rm (max)}$, respectively. Unstable circular photon orbits 
(\ref{impact1}) or (\ref{impact2}) control the shadow shapes below the red line, while the throat 
solely determines them above the blue line. In between, both effects forms the shapes 
of the shadow. Dots (a) - (h) indicate the parameters used in Fig.~\ref{fig3}.
\label{fig1}}
\end{figure}

As mentioned above, it was overlooked in Ref.~\cite{Amir:2018pcu} that the radius of the unstable 
circular photon orbit given in Eqs.(\ref{impact1}) with (\ref{rph}), or (\ref{impact2}), would be 
smaller than the throat radius $r_{\rm throat}$ for some parameter ranges. This is the case that 
the unstable circular photon orbit is located at the throat: 
\begin{equation}
r_{\rm ph} = r_{\rm throat}.
\end{equation}
The throat effect starts to appear when
$r_{\rm throat}$ becomes larger than $r_{\rm ph}^{\rm (min)}$, and completely dominates for 
$r_{\rm throat} > r_{\rm ph}^{\rm (max)}$. These regions are depicted in Fig.~\ref{fig1}.
Red and blue lines represent $r_{\rm throat}=r_{\rm ph}^{\rm (min)}$ and 
$r_{\rm throat} = r_{\rm ph}^{\rm (max)}$, respectively. The throat affects the shadow partially 
for the parameters above the red line, and determines the shadow completely for those 
above the blue line.

In this case, using, in addition, the second condition in Eq.(\ref{cond2}), which leads 
to $X(r_{\rm ph})=0$, we obtain the relation between $\eta$ and $\xi$ as
\begin{align}
\eta & = \frac{(r_{\rm throat}^2+a^2-a\xi)^2}{\Delta(r_{\rm throat})} - (\xi-a)^2 \notag \\
& = \frac{\left[ 2M(1+\lambda^2)r_{\rm throat}-a\xi\right]^2}{2M\lambda^2 r_{\rm throat}} - (\xi-a)^2,
\label{impact3}
\end{align}
where we use Eqs.(\ref{Delta-hat}) and (\ref{Delta}) in the last line. 

The third condition in Eq.(\ref{cond2}) is reduced to $X'(r_{\rm ph})>0$.
This results in 
\begin{equation}
\eta < \frac{2r_{\rm throat}}{r_{\rm throat}-M}\left[ 2M(1+\lambda^2)r_{\rm throat}-a\xi\right] - (\xi-a)^2.
\label{cond3}
\end{equation}

\section{shadows}
\label{sec_shadow}
The observer sees the projection of the unstable circular orbits, (\ref{impact1}) with (\ref{rph}), 
(\ref{impact2}), or (\ref{impact3}), as the boundary of the wormhole shadow at the so-called 
observer's sky, the plane passing through the wormhole and normal to the line of sight of the observer. 
The celestial coordinates $\alpha$ and $\beta$ measure the positions on this plane, 
and they can be written as \cite{celestial}
\begin{equation}
\begin{split}
\alpha & = \lim_{r_{\rm obs}\rightarrow\infty}\left(-r^2\sin\theta_0\frac{d\phi}{dr}\right)_{\rm obs}, \\
\beta & = \lim_{r_{\rm obs}\rightarrow\infty}\left(r^2\frac{d\theta}{dr}\right)_{\rm obs},
\end{split}
\end{equation}
where $\theta_0$ denotes the inclination angle between the rotation axis of the wormhole and the line of
sight, and the subscripts ``obs'' represent the values evaluated at the observer's position. 
Using the geodesic equations (\ref{phi_eq}) - (\ref{theta_eq}), we have \cite{Amir:2018pcu}
\begin{equation}
\begin{split}
\alpha & = -\frac{\xi}{\sin\theta_0} = -\xi, \\
\beta & = \pm \sqrt{\eta+a^2\cos^2\theta_0-\frac{\xi^2}{\tan^2\theta_0}} = \pm\sqrt{\eta},
\end{split}
\label{cel_cor}
\end{equation}
where the last equalities are estimated when the observer is located on the equatorial plane 
($\theta_0=90^\circ$).

\subsection{Non-rotating case ($a=0$)}
\begin{figure}[ht!]
\includegraphics[width=90mm]{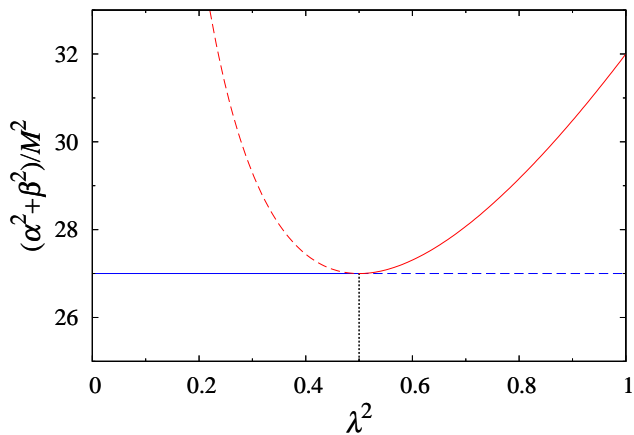} 
\caption{Radius of the shadow on the celestial plane for $a=0$, shown in Eq.(\ref{a2b2}). 
Red and blue lines represent respectively the cases with and without the effects of the throat. 
Solid lines express the actual sizes of the shadows.
\label{fig2}}
\end{figure}

It is instructive to show the non-rotating case ($a=0$) in the first place. In this case,
$r_{\rm ph} = 3M $ for $\lambda^2 \le \frac{1}{2}$, while $r_{\rm ph}=r_{\rm throat}=2M(1+\lambda^2)$ 
for $\lambda^2 > \frac{1}{2}$, and the shadow exhibits just a circle on the celestial plane as 
\begin{equation}
\alpha^2+\beta^2 = 
\begin{cases}
27 M^2, & (\lambda^2 \le \frac{1}{2}), \\[1mm]
\displaystyle{\frac{4(1+\lambda^2)^3}{\lambda^2}M^2}, &  (\lambda^2 > \frac{1}{2}).
\end{cases}
\label{a2b2}
\end{equation}
This is shown in Fig.~\ref{fig2}. Red and blue lines denote the cases for $\lambda^2 > \frac{1}{2}$ and
$\lambda^2 \le \frac{1}{2}$, respectively. The solid lines show the actual radius of the shadow. Therefore,
the size of the shadow becomes larger for $\lambda^2 > \frac{1}{2}$, which is
the impact of the throat.

Notice that $\lambda=0$ reduces to the non-rotating black hole case whose shadow radius is
estimated as $\alpha^2+\beta^2 =27M^2$, the same as that of the Schwarzschild-like wormhole 
for $0<\lambda^2 \le \frac{1}{2}$.

The third condition in Eq.(\ref{cond2}) is rewritten as
\begin{equation}
\alpha^2 +\beta^2 < 
\begin{cases}
\displaystyle{54 M^2}, & (\lambda^2 \le \frac{1}{2}), \\[4mm]
\displaystyle{\frac{16(1+\lambda^2)^3}{1+2\lambda^2}M^2}, &  (\lambda^2 > \frac{1}{2}),
\end{cases}
\end{equation}
where Eqs.(\ref{cond3-0}) and (\ref{cond3}) are used. Therefore, $\alpha^2+\beta^2$ in Eq.(\ref{a2b2})
always meets this condition.\footnote{
Similar argument on unstable circular photon orbits was given for the $a=0$ 
case in Ref.~\cite{Tsukamoto:2020uay}.}

\subsection{Rotating case ($a\ne0$)}
Now let us move on to the general situations when the wormhole is rotating ($a\ne0$). We 
consider the case when the observer is on the equatorial plane ($\theta_0=90^\circ$), since
it is the most apparent situation to see the effect of rotation. 

\begin{figure*}[ht!]
\begin{tabular}{cc}
\includegraphics[width=80mm]{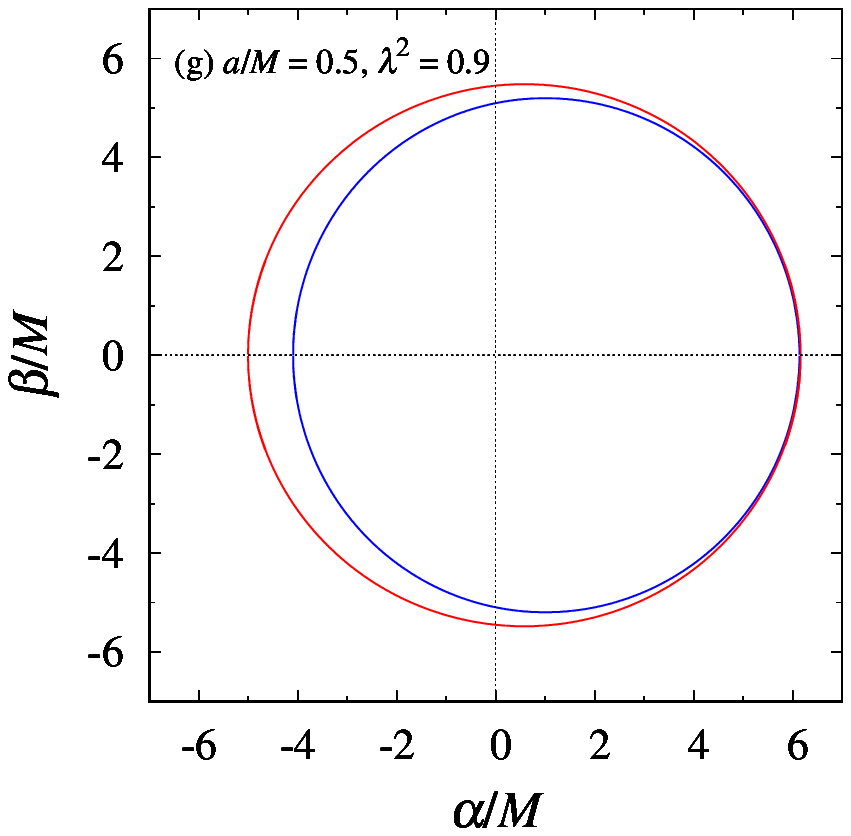} &
\includegraphics[width=80mm]{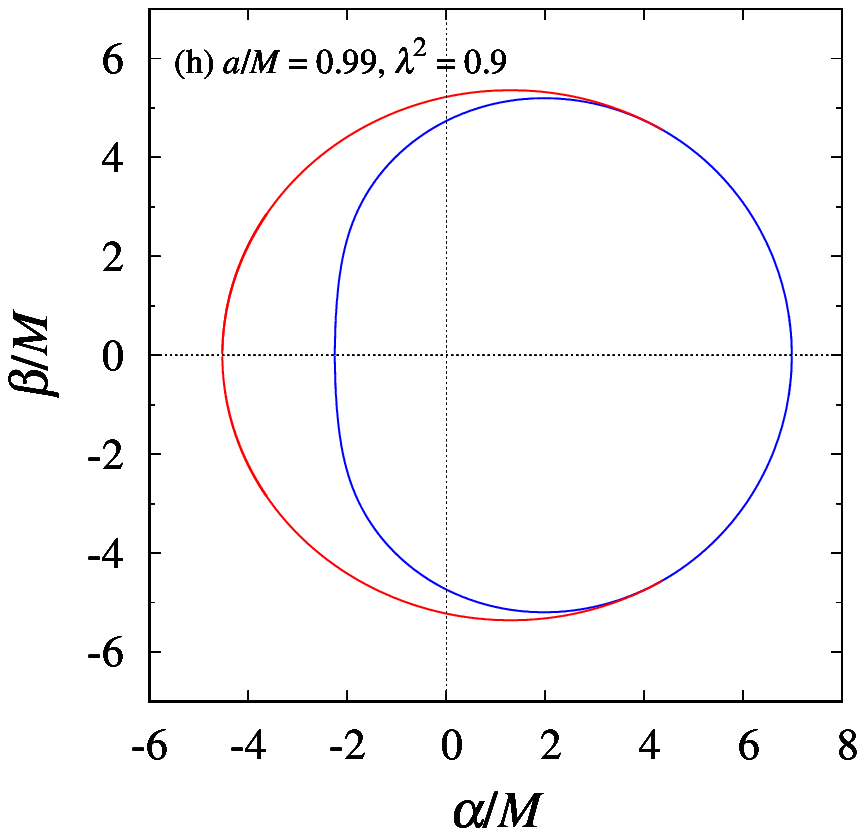} \\
\includegraphics[width=80mm]{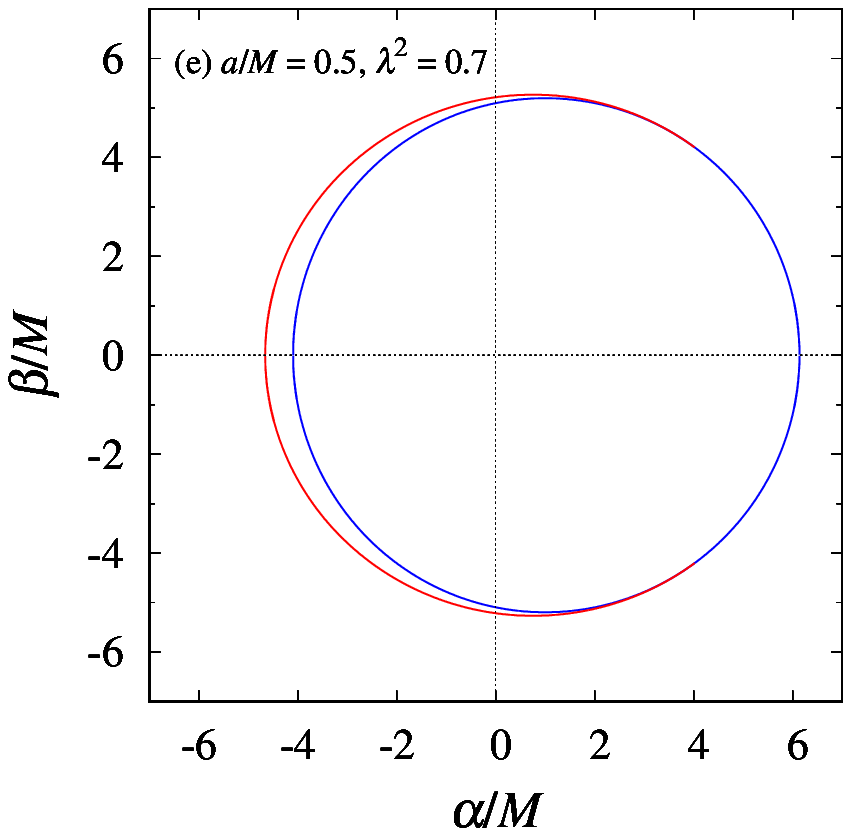} &
\includegraphics[width=80mm]{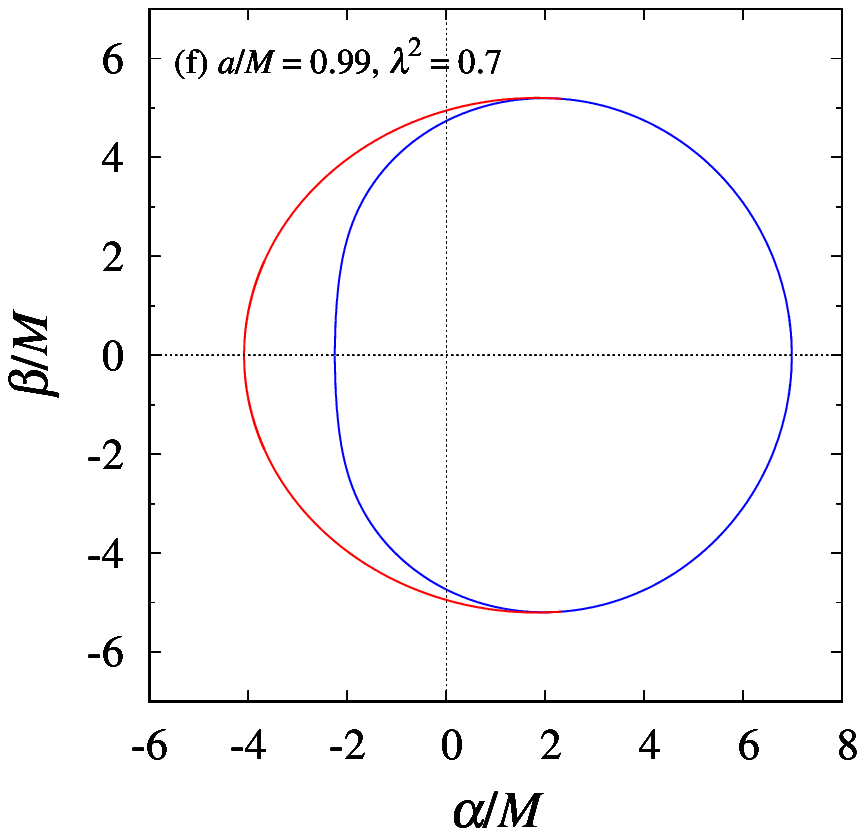} \\
\includegraphics[width=80mm]{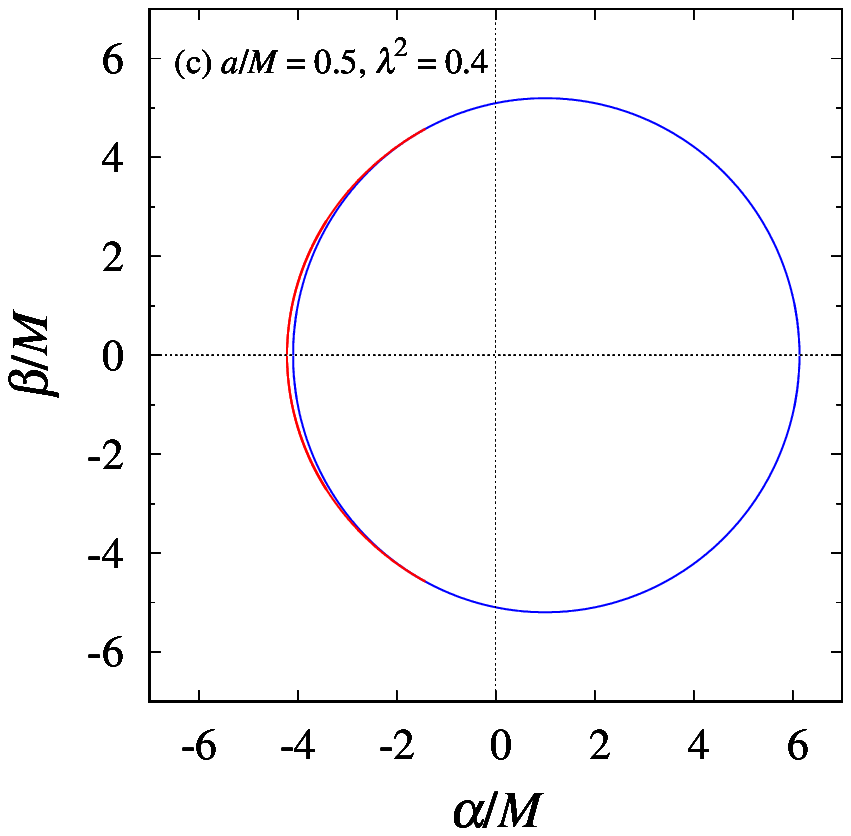} &
\includegraphics[width=80mm]{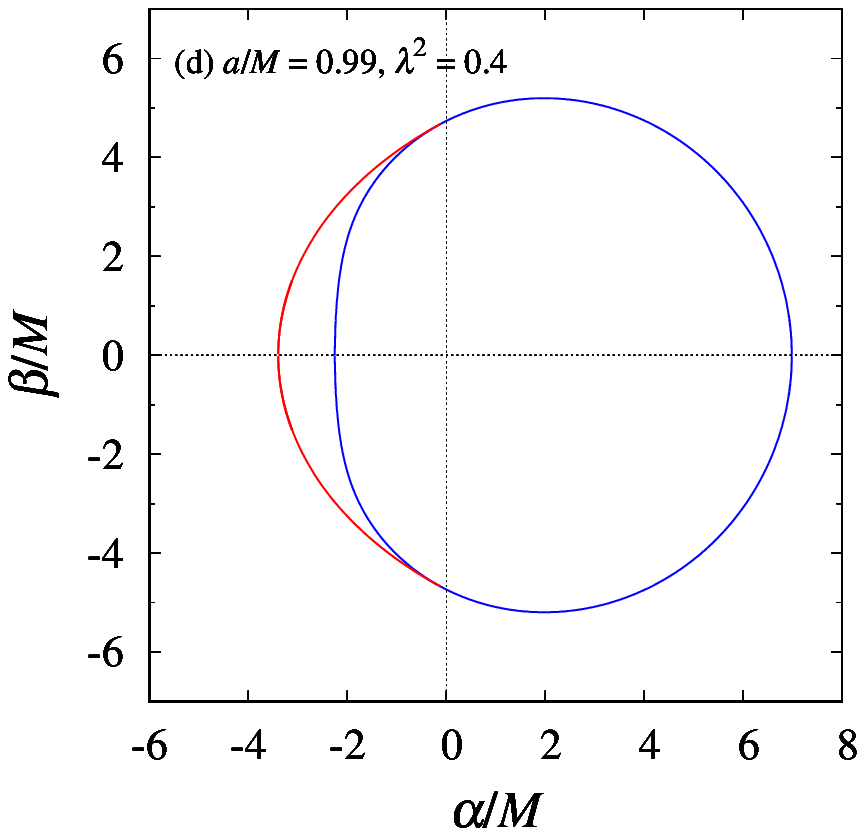} \\
\includegraphics[width=80mm]{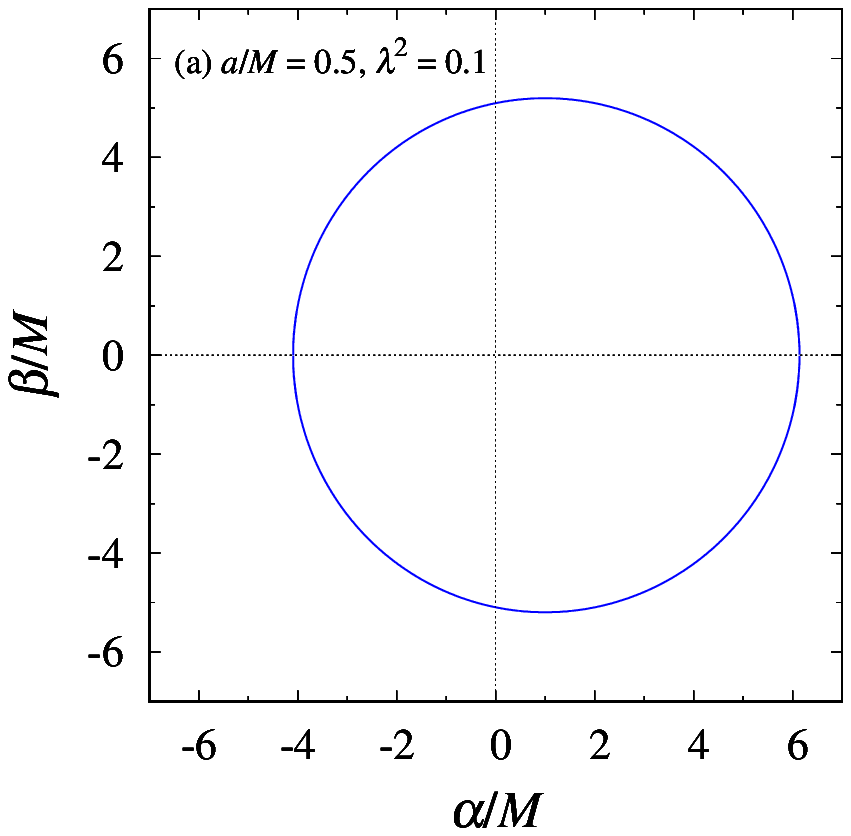} &
\includegraphics[width=80mm]{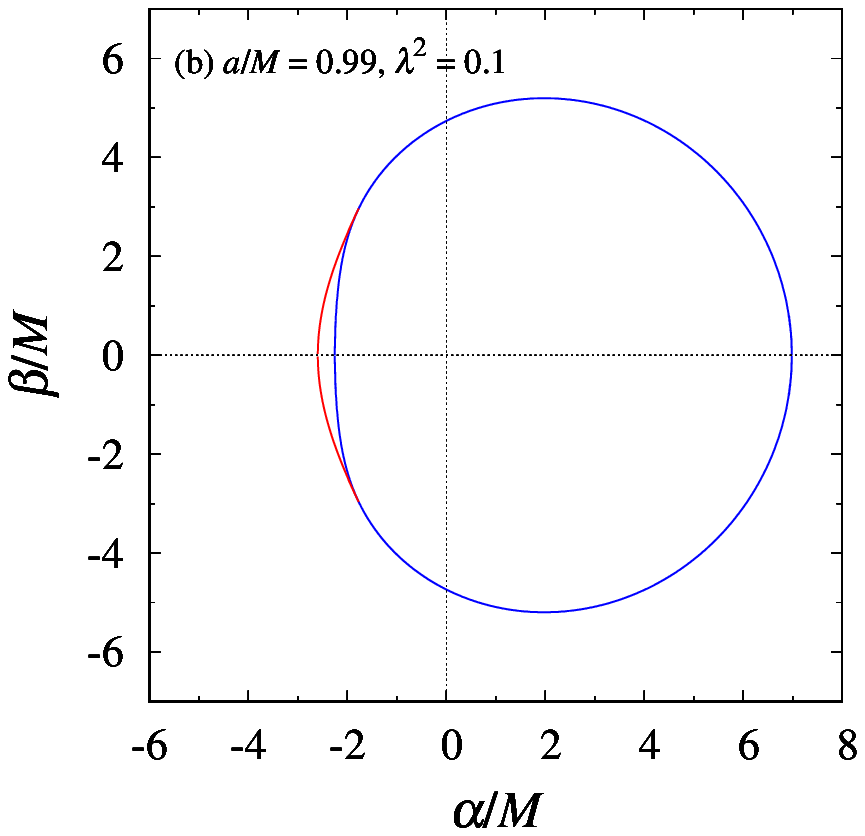}
\end{tabular}
\caption{Shadows of the Kerr-like wormholes on the celestial plane for $a=0.5$ in the left row and 
$a=0.99$ in the right row with $\lambda^2=0.1$, 0.4, 0.7, and 0.9 from the bottom to the top, respectively.
(a) - (h) cases correspond to the parameters marked with dots in Fig.~\ref{fig1}.
Shadows of the wormhole without throat effects is displayed 
by solid blue lines, which is exactly the same as the shadows of the Kerr black hole.
Red lines come from the effects of the wormhole throat.
\label{fig3}}
\end{figure*}

In order to find the shape of the shadow, we should insert the impact parameters (\ref{impact1}) 
into the celestial coordinate (\ref{cel_cor}), and vary the unstable circular photon radius 
$r_{\rm ph}$ in the range (\ref{rph}). This is the case which the throat has nothing to do with,
and we show the shadow for $a=0.5$ and $\lambda^2=0.1$ in solid blue line in Fig.~\ref{fig3}(a) as an example. 
Note that the shape of the shadow is the same for the Kerr black hole with the same angular momentum 
and the same mass as the Kerr-like wormhole.

The effects of the throat on the shadow appear for larger $a$, which was overlooked
in Ref.~\cite{Amir:2018pcu}. We show an example case with $a=0.99$ and $\lambda^2=0.1$ in 
Fig.~\ref{fig3}(b). Here again the shadow of the wormhole without throat effects is displayed 
by a solid blue line.

Red line comes newly from the effects of the wormhole throat considered in Sec.~\ref{sec_throat}. 
Here, we put the impact parameters in Eq.(\ref{impact3}) into celestial coordinates (\ref{cel_cor}) 
to find the boundary on the celestial plane. Therefore, the shape of the wormhole shadow
does not shrink too much for the prograde orbits because of the existence of the wormhole
throat. As a result, the red line deforms the shape of the wormhole shadow so that we can distinguish
between the shadows of a wormhole and a black hole.

The effects of the throat becomes more remarkable in larger $\lambda$ cases. 
We show these cases in Fig.~\ref{fig3}(c)-(f) and (h), where the throat effects
deform the shapes of the shadows to certain extents. In these cases, the radius of
the throat $r_{\rm throat}$ is larger than the prograde radius $r_{\rm ph}^{\rm (min)}$,
but smaller than the retrograde radius $r_{\rm ph}^{\rm (max)}$.

Finally, the whole shadow shape is determined by the throat, for example, in the case 
with $a=0.5$ and $\lambda^2=0.9$, shown in Fig.~\ref{fig3}(g). In this situation, the shadow shape 
of the wormhole in red line is completely different from that of the black hole and the wormhole without
the throat effects taken into account in blue line.
Note that the shape differs also from a circle. See Eq.(\ref{impact3}).

Note that the third condition in Eq.(\ref{cond2}) is satisfied in all the cases. 
Therefore, the obtained circular photon orbits are indeed unstable.

\section{Conclusions}
\label{concl}
We have revisited to investigate the shadows cast by the Kerr-like wormhole. The boundary of 
the shadow is determined by unstable circular photon orbits. We have found that, in certain 
parameter regions, the orbit is located at the throat of the Kerr-like wormhole, which was not considered 
in the literatures. These cases take place for larger wormhole spin $a$ and/or larger deviation 
parameter $\lambda$. The existence of the throat alters the shape of the shadow considerably.
We can thus differentiate it from that of the Kerr black hole, which is exactly the same
as that of the Kerr-like wormhole without throat effects taken into account.

On the other hand, we can get another perspective. If there are throat effects on the shadow shape,
one can figure out more easily whether the observed shadow is cast by the Kerr-like wormhole 
or the Kerr black hole even in the case when the object mass is unknown, not determined by other 
observations such as by motions of objects around the wormhole/black hole.

\section*{Acknowledgments}
The authors are grateful to Michiyasu Nagasawa and Naoki Tsukamoto for useful comments.

\appendix

\section*{APPENDIX}
For completeness, we also consider the cases when we regard the ADM mass as the wormhole mass, 
$M_{\rm WM}^{\rm (ADM)}=M(1+\lambda^2)$. Then, it is better to normalize the parameters with respect to 
$M_{\rm WH}^{\rm (ADM)}$, which implies that we set $M_{\rm WH}^{\rm (ADM)}=M_{\rm BH}^{\rm (ADM)}=1$, 
as done in Ref.~\cite{Amir:2018pcu}, where they found that the shadow radius of the Kerr-like 
wormhole becomes smaller.

\begin{figure}[ht!]
\includegraphics[width=90mm]{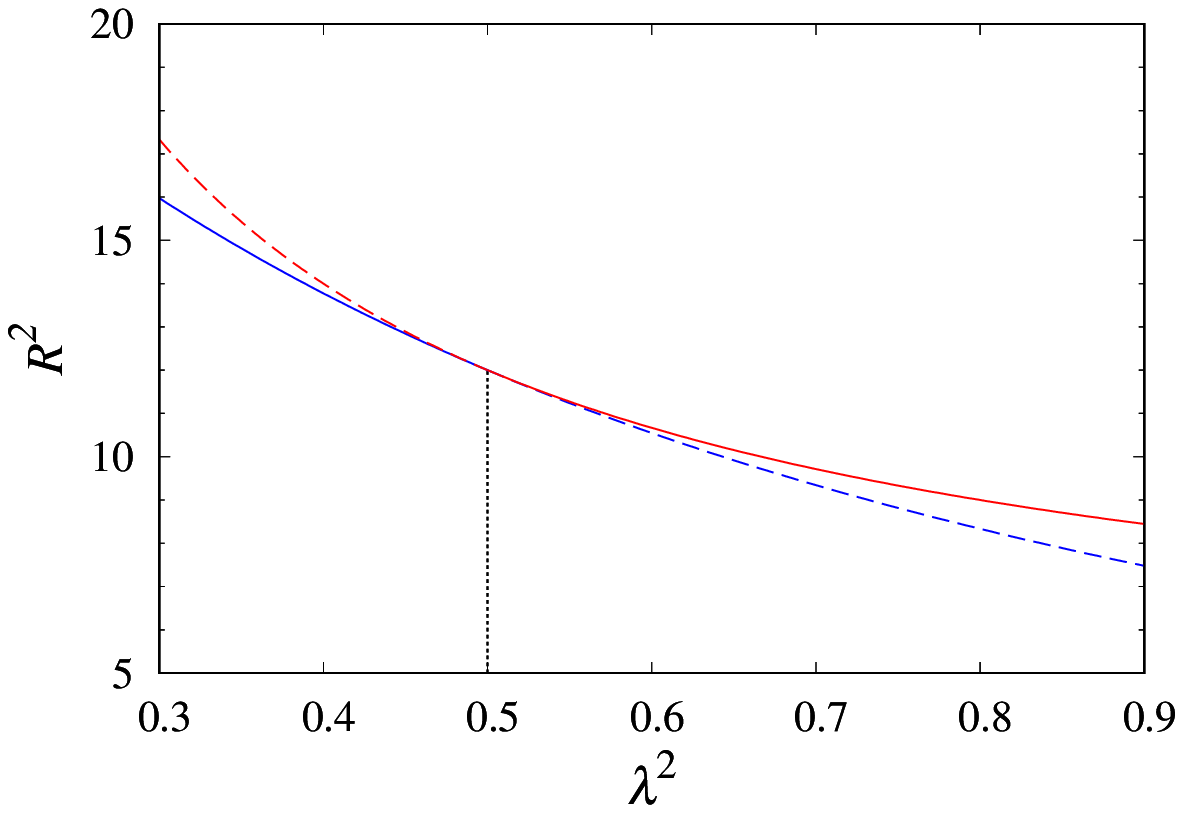} 
\caption{Radius of the shadow on the celestial plane for $a=0$, shown in Eq.(\ref{radius}). 
Red and blue lines represent respectively the cases with and without the effects of the throat. 
Solid lines express the actual sizes of the shadows.
Here we set $M_{\rm WH}^{\rm (ADM)}=1$.
\label{fig4}}
\end{figure}

For non-rotating case, the shadow radius $R$ on the celestial plane is written as
\begin{equation}
R^2 = 
\begin{cases}
\displaystyle{\frac{27}{(1+\lambda^2)^2}}, & (\lambda^2 \le \frac{1}{2}), \\[4mm]
\displaystyle{\frac{4(1+\lambda^2)}{\lambda^2}}, &  (\lambda^2 > \frac{1}{2}),
\end{cases}
\label{radius}
\end{equation}
displayed in Fig.~\ref{fig4}. $\lambda=0$ reduces to the non-rotating black hole case whose 
shadow radius is $R^2 =27$. 

\begin{figure*}[ht!]
\begin{tabular}{cc}
\includegraphics[width=80mm]{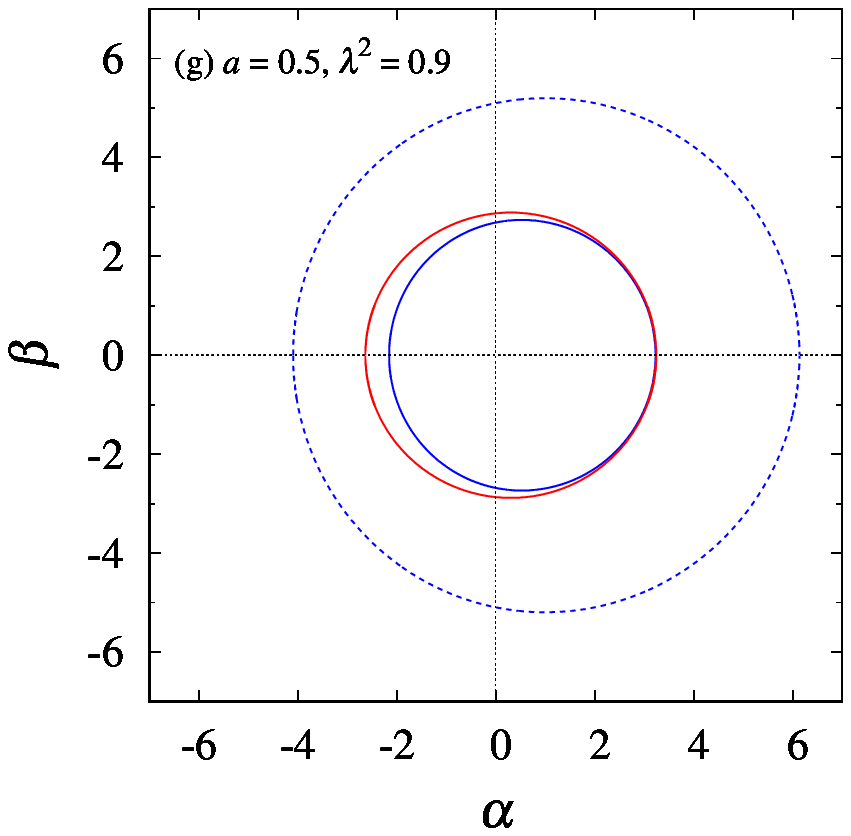} &
\includegraphics[width=80mm]{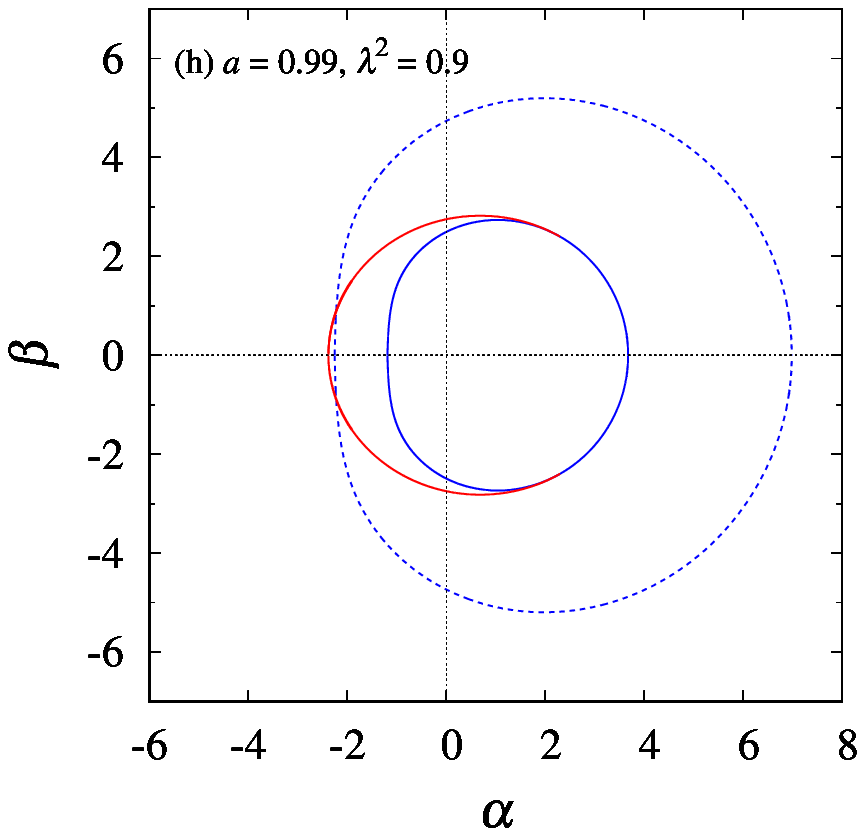} \\
\includegraphics[width=80mm]{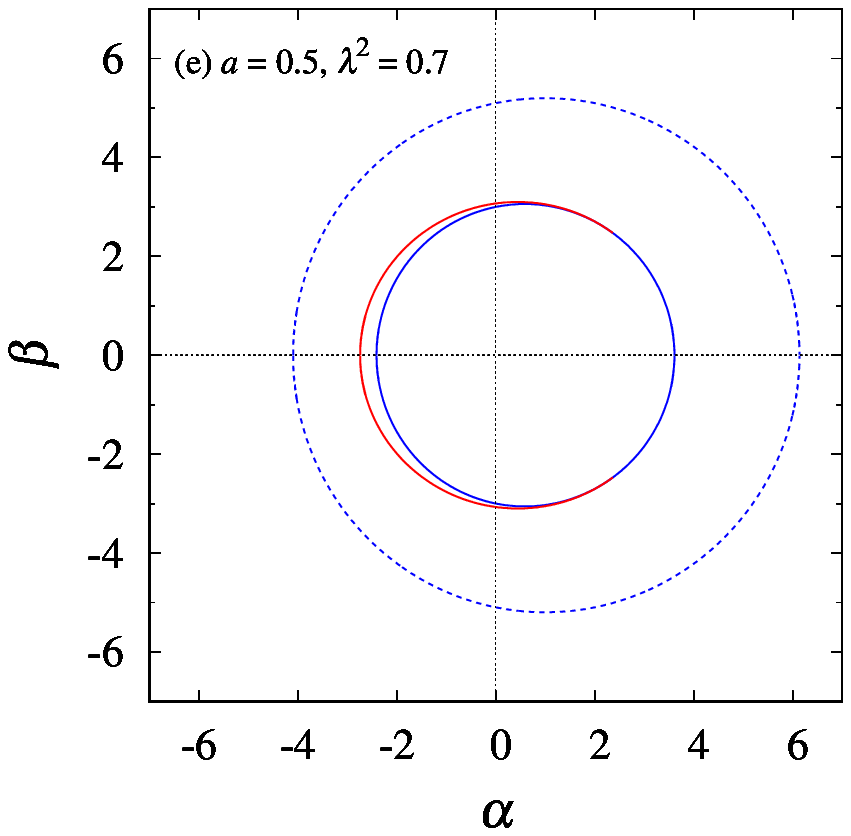} &
\includegraphics[width=80mm]{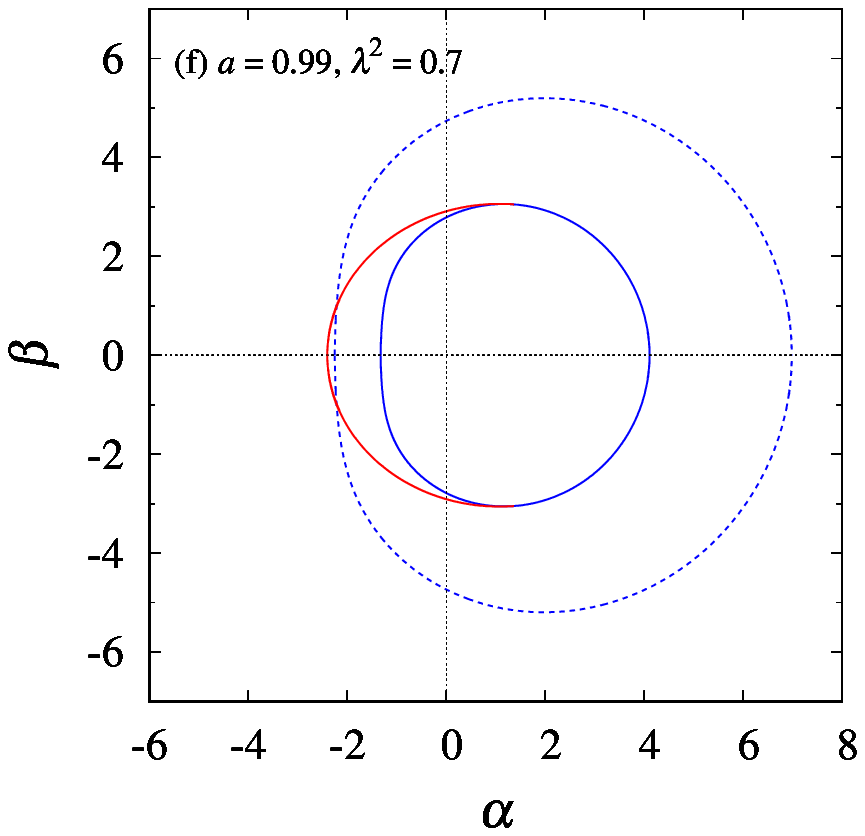} \\
\includegraphics[width=80mm]{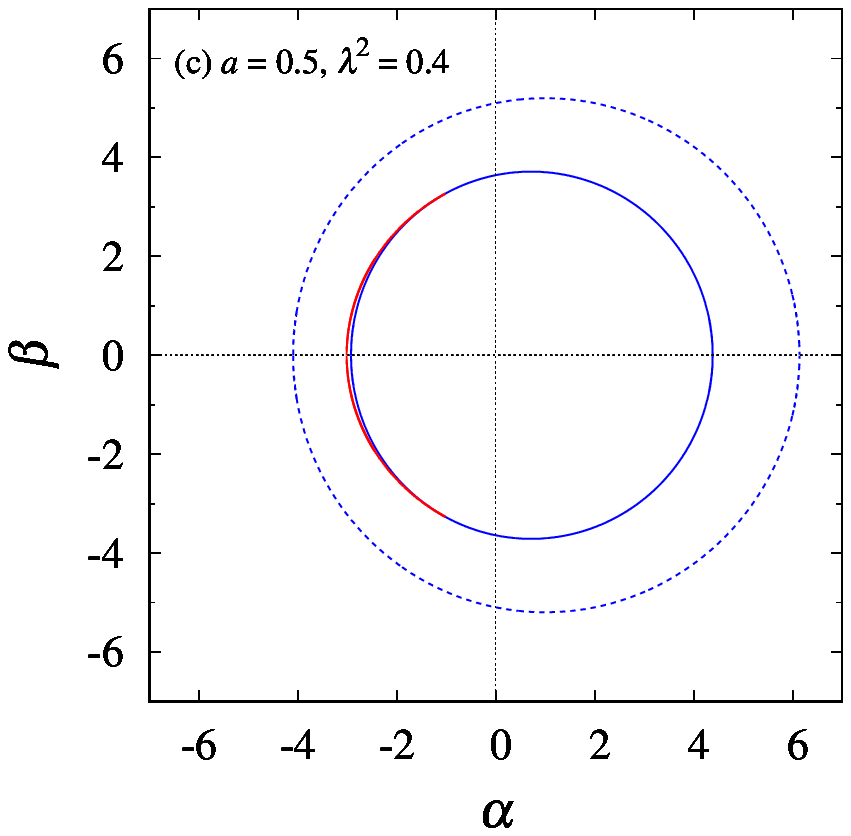} &
\includegraphics[width=80mm]{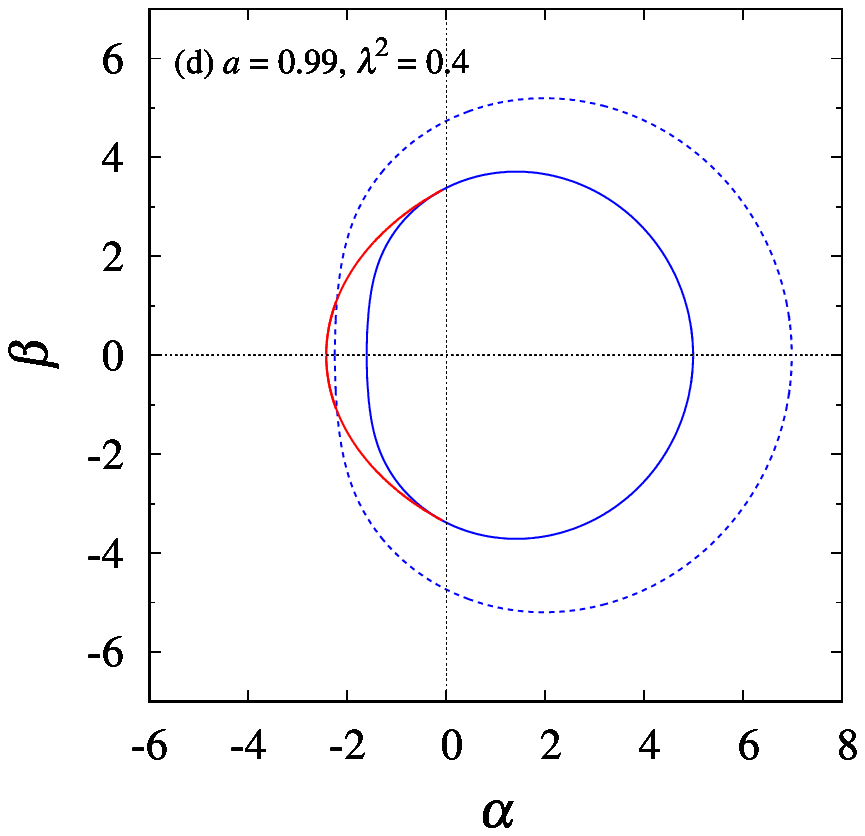} \\
\includegraphics[width=80mm]{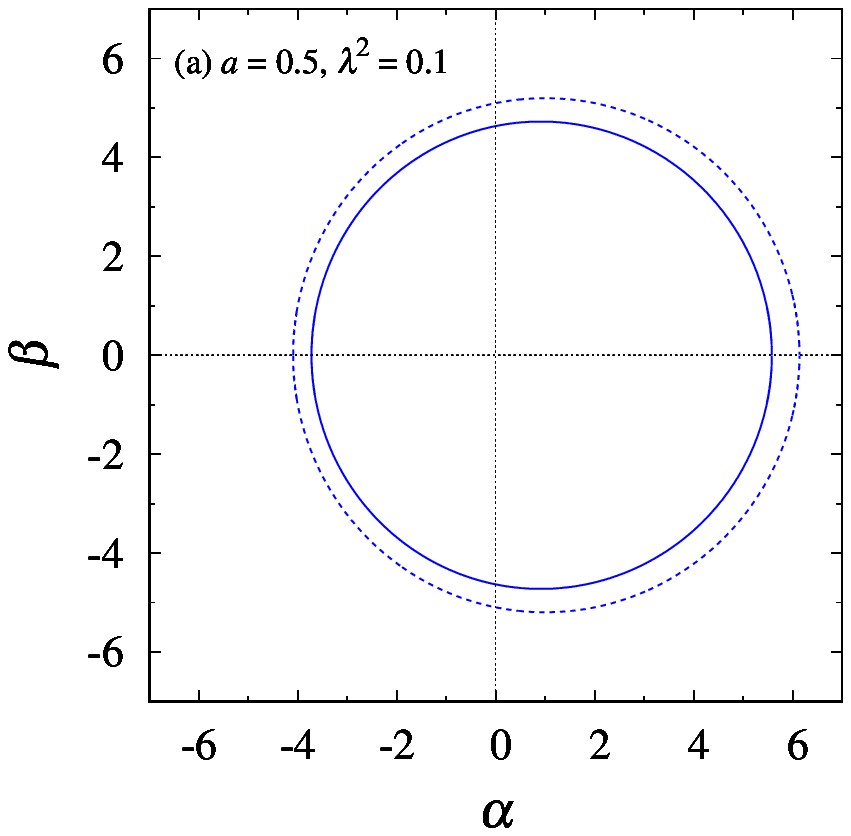} &
\includegraphics[width=80mm]{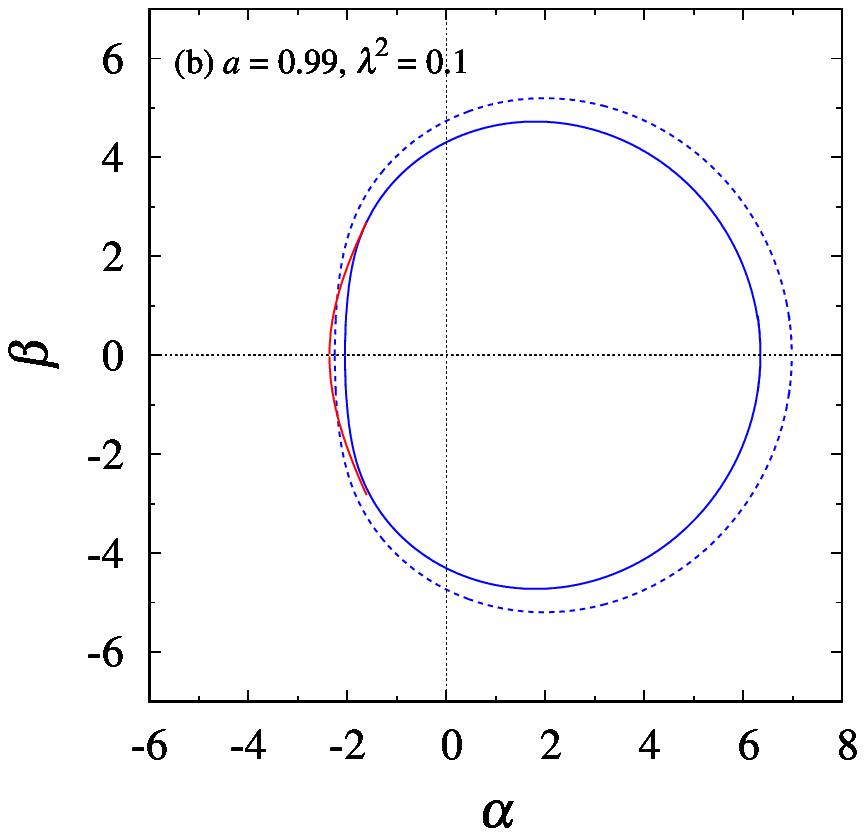}
\end{tabular}
\caption{Shadows of the Kerr-like wormholes on the celestial plane for $a=0.5$ in the left row and 
$a=0.99$ in the right row with $\lambda^2=0.1$, 0.4, 0.7, and 0.9 from the bottom to the top, respectively.
(a) - (h) cases correspond to the parameters marked with dots in Fig.~\ref{fig1}.
Shadows of the wormhole without throat effects is displayed 
by solid blue lines, while we denote the shadows of the black hole in dashed blue lines.
Red lines come from the effects of the wormhole throat. 
Here we set $M_{\rm WH}^{\rm (ADM)}=M_{\rm BH}^{\rm (ADM)}=1$.
\label{fig5}}
\end{figure*}

On the other hand, we show rotating cases in Fig.~\ref{fig5} for the same $a$ and $\lambda$ as in 
Fig.~\ref{fig3}. We can see that the throat effects alter the shadow shapes more 
considerably for larger $a$ and $\lambda$.

\end{document}